# Mapping parcel-level urban areas for a large geographical area

**Ying LONG**
Beijing Institute of City Planning, Beijing, China
Department of Architecture, University of Cambridge, UK
Email: longying1980@gmail.com

**Yao SHEN**
The Bartlett, University College London, UK
Email: y.shen.12@ucl.ac.uk

## ABSTRACT

As a vital indicator for measuring urban development, urban areas are expected to be identified explicitly and conveniently with widely available dataset thereby benefiting the planning decisions and relevant urban studies. Existing approaches to identify urban areas normally based on mid-resolution sensing dataset, socioeconomic information (e.g. population density) generally associate with low-resolution in space, e.g. cells with several square kilometers or even larger towns/wards. Yet, few of them pay attention to defining urban areas with micro data in a fine-scaled manner with large extend scale by incorporating the morphological and functional characteristics. This paper investigates an automated framework to delineate urban areas in the parcel level, using increasingly available ordnance surveys for generating all parcels (or geo-units) and ubiquitous points of interest (POIs) for inferring density of each parcel. A vector cellular automata model was adopted for identifying urban parcels from all generated parcels, taking into account density, neighborhood condition, and other spatial variables of each parcel. We applied this approach for mapping urban areas of all 654 Chinese cities and compared them with those interpreted from mid-resolution remote sensing images and inferred by population density and road intersections. Our proposed framework is proved to be more straight-forward, time-saving and fine-scaled, compared with other existing ones, and reclaim the need for consistency, efficiency and availability in defining urban areas with well-consideration of omnipresent spatial and functional factors across cities.

## ACKNOWLEDGMENTS

The first author acknowledges the financial support of the National Natural Science Foundation of China (No.51078213 and 51311120081) and China Scholarship Council (No.201209110066). Our thanks also go to Dr Kang Wu and Dong Li for providing China datasets. We would release the mapping results online on acceptance of this paper.

# 1 INTRODUCTION

This study is a manual for making a 'parcel-up' understanding of urban areas for a larger geographical area: urban areas here are defined as the mergence of urban parcels enclosed by roads with more urban activities. Dissecting the existing body of influential approaches to define urban areas and examining cases, it identifies a link between two dominant factors that are instrumental in achieving the goal. The first is the urban parcel demarcated by the roads, which could be considered as the basic unit of urban areas; the second is urban density, which is the key function characteristic, manifested in POIs density.

A universal difficulty for urban studies is how a city can be defined properly (Zipf, 1949; Krugman 1996; Soo, 2005; Batty, 2006). Urban built-up areas (urban areas in the following context) play a strong role in representing urban spatial development for planning decisions, management, and urban studies. They not only illustrate spatial patterns, such as the development levels and scales of the built environment, but also reveal socio-economic unevenness within the built-up areas, e.g., population aggregation, social interaction energy consumption, and land use efficiency, thereby representing how a city evolves in a complex manner (Batty, 2011). Conventional methods of capturing the borders of a built-up area from the top down have been applied in major cities around the world on a large scale mainly relying on mid-resolution sensing dataset or socioeconomic distributions (e.g. population density) associating with low-resolution, e.g. cells with several square kilometers or even larger towns/wards. However, such methods cannot be applied to most of cities in developing countries due to lacking necessary data (Long and Liu, 2013). Moreover, these existing methods still require multiple steps according to unique conditions if achieving a fine scale result is expected. This paper develops an automated approach for producing fine-scaled urban areas of all different-sized Chinese cities based on morphological and functional characteristics determined by road network layer from ordnance survey and points of interest (POIs).

Urban area is a widely applied, discussed and referred concept but ambiguous simultaneously. In existing literature, it shows various descriptions, measurements and applications spanning various issues and spatial scales in different nations. Urban areas in US are identified as Urbanized Areas (UA) in a typical administrative model for spatial statistics containing the incorporated places and census designated places in central places and urban fringes controlling for the population density (Morrill et al., 1999). One similar term in Japan is Densely Inhabited District (DID) with population density over 4000 people per $Km^2$. In China, the records of 'one book and two certificates'[1] within administrative areas are widely accepted (Xu and Hua, 2005). Furthermore, Urban Areas (UA) in UK are derived from constructions-built areas where certain real-estate densities are detected through satellite images or other datasets (Hu et al., 2008). On the other hand, socio-economic factors are also adopted to describe the active urban areas, e.g. labor force markets and commuter sheds are utilized to represent Metropolitan Areas (MA) in US (Berry, et, at., 1969). Despite definitions of urban built-up areas are highly ambiguous, well identified urban areas needs explicitly bonding the spatial and functional dimensions. In this sense, urban areas in this study are further understood as the urban extent containing urban parcels with actual usage which is reflected by the POIs.

Parallel to the ambiguity of its definitions, there are various distinguished methods for mapping urban areas. From the morphological perspective, remote sensing images and road network have received increasing academic attention. Remote sensing or night-time satellite images help to filter non-built areas on the basis of transferring land cover information or scanned light brightness to indices (Henderson et. al, 2003, He et al., 2006). In addition, various geometrical characteristics of road networks have been introduced to identify the spatial organization of cities as physical entities, e.g. road intersection density (Masucci et al., 2012); fractal indices (Shen, 2002; Tannier and Tomas 2011,2013; Jiang and Yin, 2013); size of urban blocks (Jiang and Liu, 2012). In terms of the functional aspects, applying socio-economic statistics such as demographic densities (Rozenfeld et al., 2008), effective employment density (SGS Economics & Planning, 2011, 2012) and infrastructures accessibilities (Hu et al., 2008) has emerged as a standard method of defining urban areas. However, In case more precise outcomes with higher-resolution are expected, the disadvantages of these approaches are not rare as follows. 1) remote sensing data based approaches are limited by the time-consuming interpretation steps and the image resolution; 2) to accurately define cities, using the geometrical approach to directly link to specific spatial units, such as a parcel, block, or tile, is difficult despite of its advantage in cities with diverse sizes at an extremely large scale; 3) spatial statistics methods are very time consuming to prepare, limited by fine-scaled censuses, and the results are most likely to be altered significantly if the low survey frequency.

---

[1] "One book and two certificates" - containing proposal of project location, permit of land planning and permit of construction planning - refers to the construction approval files delivered by the government based on urban planning law.

Hence, many challenges need to be overcome before a universal model can be established. The foremost challenge is the question of how fine-scaled spatial units can be set and which one will be suitable for urban studies and planning practice. Furthermore, the data quality for different cities might not permit the development of a universal approach. In addition, a difficult task is finding a straightforward way to generate an urban indicator with consideration of various socio-economic aspects.

Recently, these challenges of urban area delineation gave rise to methodological developments that address the same issue from the bottom up based on detailed street network and Volunteered Geographic Information. Several studies concentrated on extracting a parcel-based urban area from the transport layer in OpenStreetMap (OSM) (Jiang and Liu, 2012; Jiang, et al., 2013). Yet, pure road network-based approach is hardly effective for generating fine-grained parcels and inferring urban parcels using the head-tail division rule (globally applied) to reflect real urban activities. With this background taken into account, some studies have been conducted to utilize POIs for inferring the function performance of auto-generated parcels so that the urban parcels could be selected locally. Yuan et al. (2012) segmented Beijing into disjointed parcels through the raster-based model, and their functional characteristics were inferred by incorporating the POIs and the taxi trajectories. Long and Liu (2013) proposed an approach for automatically identifying and characterizing parcels (AICP) by using OSM and POIs in 297 Chinese cities at the national scale. They compared the efficiency and accuracy of the approach with those of other methods. Apart from previous research, these two studies shed light not only on the auto-generation of parcels but also on the functional qualification in terms of the online volunteered data e.g., POIs. However, the unevenness of the resolution of OSM among various cities (Hagenauer and Helbich, 2012) still limits the applicability of these methodologies for all cities and the resolution of the results.

Compared with the quality of OSM, the ordnance survey map - a national-scale authoritative dataset - maintains the best completeness and coverage in all cities. Within the context of Web 2.0, the commercial ordnance survey data became accessible to the public in developing countries in recent years, which enabled all Chinese cities (big or small) to be included thus mitigating the digital divide existed before. High correspondence between commercial ordnance road layer and POI locations in the datasets of navigation firms helped us to combine the spatial and functional factors at the same resolution for all Chinese cities. Few relevant studies used these two datasets simultaneously to produce fine-scaled urban areas for all the cities.

This work adopts the framework of AICP proposed by Long and Liu (2013) with the ordnance survey and POIs for inferring urban areas at the parcel level for all Chinese cities. A comparison between the road network in the survey and that in OSM indicates that the ordnance survey map produced more detailed information especially in medium and small cities. In response, road network in the ordnance survey is imposed into the AICP model to generate more realistic urban parcels, which are the basic geographic units to describe urban areas of the cities. When the produced urban areas are overlapped with the ones generated based on other datasets, the results indicate that the extracted urban areas based on the ordnance survey and POIs are significantly more accurate in middle and small-sized cities than in major cities. Our findings also indicate that AICP could be utilized as an open and direct approach by inputting high-resolution and ubiquitous data to capture small parcel-based urban areas. This study aims to offer an alternative way to understand complex urban systems across cities from the bottom up. This paper is structured as follows. Section 2 describes the datasets used in this paper. The methods and their results are introduced in Section 3 and 4. Section 5 and 6 discuss the results and make concluding remarks of this research, respectively.

## 2 DATA

### 2.1 Administrative boundaries of Chinese cities

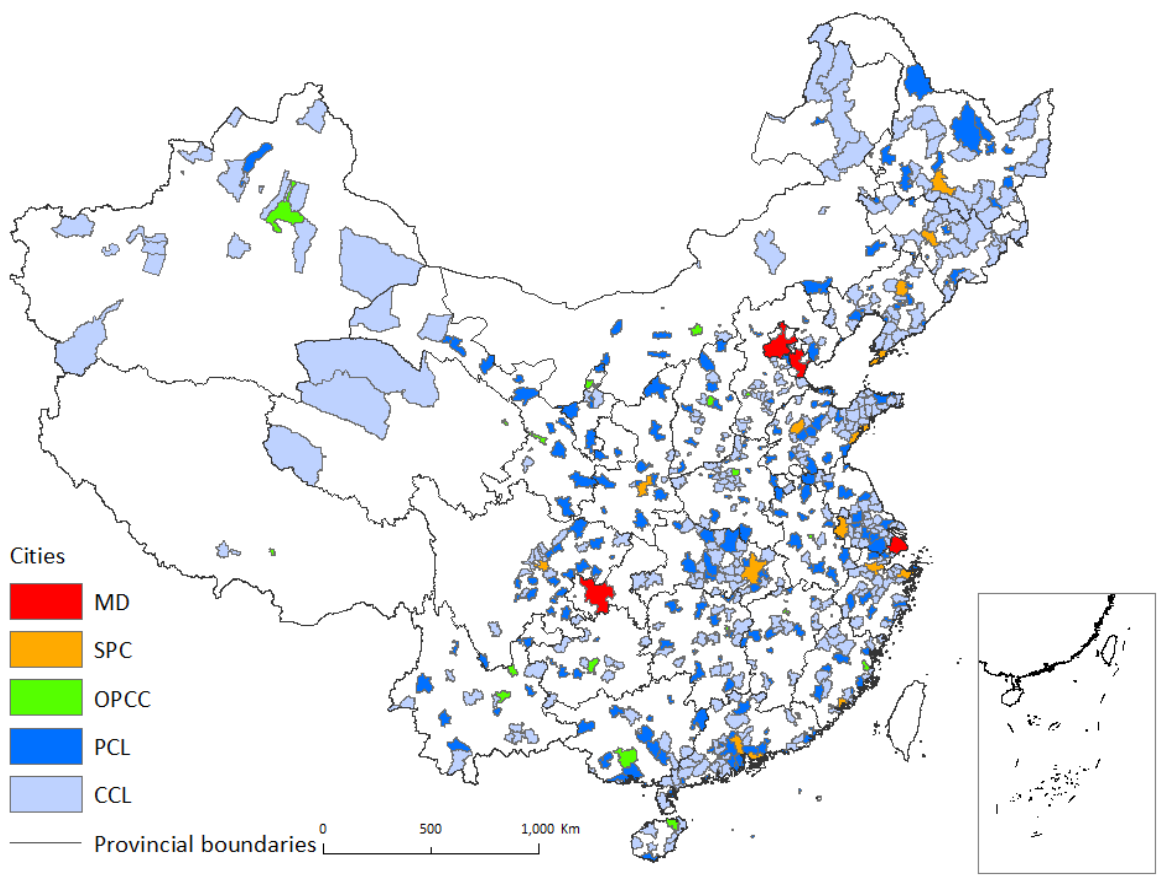

Figure1 Administrative areas of Chinese cities

Administrative boundaries of 654 Chinese cities[2] - the limitations of local geographies - are applied to partition the whole research areas to legal cities so that ordnance survey maps and POIs can be curved off accordingly (Figure 1). According to Chinese administrative system (Ministry of Housing and Urban Development, 2012; Ma, 2005), there are mainly five levels of cities classified in this way including: municipalities directly leaded by the nation (MD, 4 cities), sub-provincial cities (SPC, 15 cities), other provincial cities (OPCC, 17 cities), prefecture-level cities (PLC, 250 cities), and county-level cities (CLC, 368). This system can also reflect hierarchy of these cities in terms of city size and population. By doing so, the research scopes are specified to administrative areas of urban lands by keeping the national scale in mind thereby well controlling the local situation of each city from a global perspective of the whole nation.

---

[2] Sansha in Hainan and Beitun in Xinjiang appearing in MOHURD (2013) were not included due to spatial data availability. Taiwan was not included in all analysis and results in this paper.

## 2.2 Total urban area of Chinese cities

Based on defined city administrative boundaries, statistics of urban area are extracted from MOHURD (2013) in order to allocate the total into urban parcels in each city. Until 2012, total urban area of 654 cities in China has reached 46,744 km$^2$. Individual city is inferred by its statistical area decently (Figure 2). Consequently, our research areas in all the cities are specifically featured with their administrative subordination and total urban area.

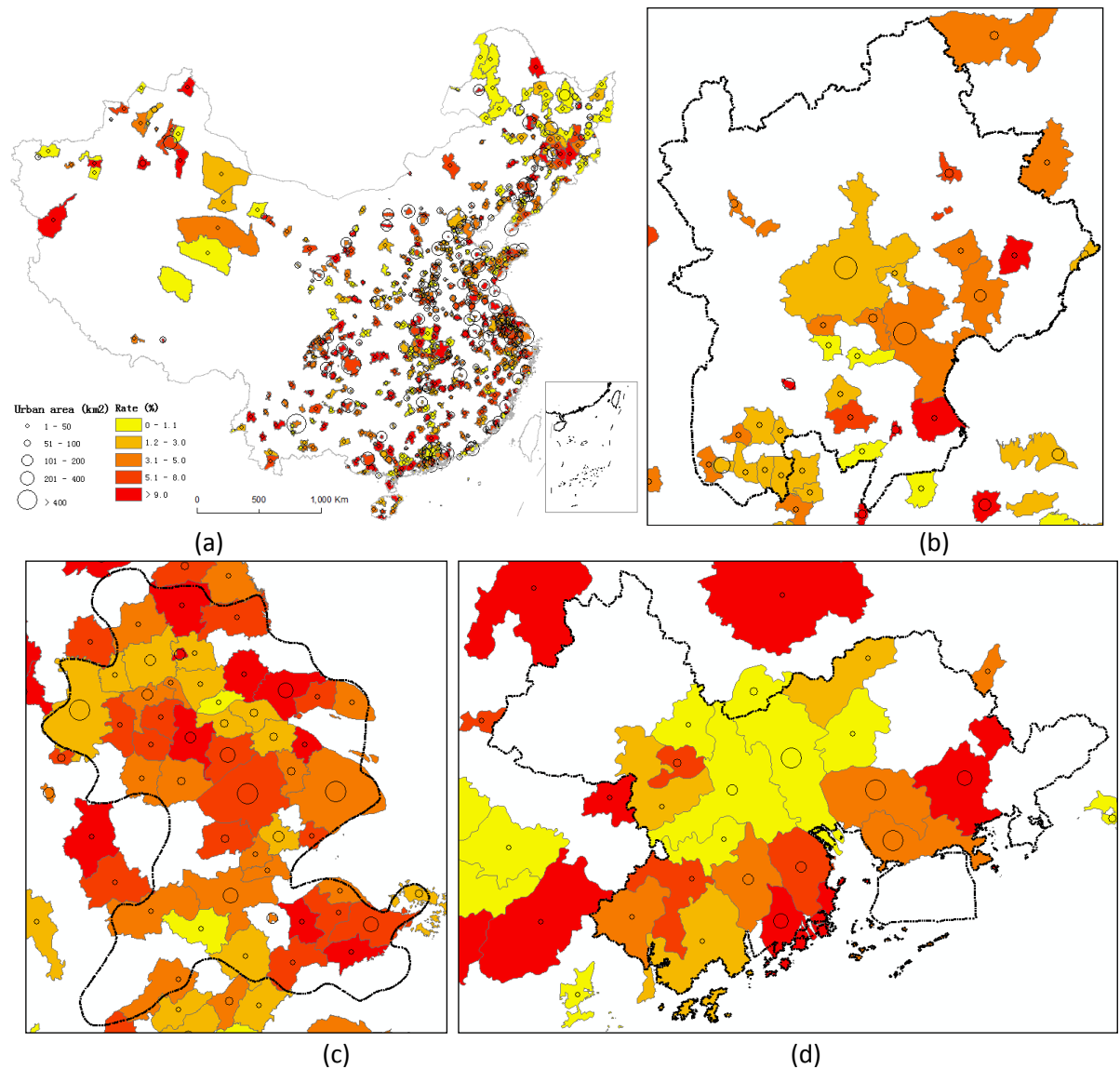

Figure 2 Total urban area in 2012 at the city level. (a) the whole China, (b) Beijing-Tianjin-Hebei (BTH), (c) Yangtze River Delta (YRD), (d) Pearl River Delta (PRD)
Note: The urban expansion rate during 2007-2012 of each city is also mapped in this figure to show the historical urban expansion of Chinese cities.

## 2.3 Road network in ordnance survey and POIs in 2012

The ordnance map is considered to be the authorized map reflecting the most urban information (Haklay, 2010). Urban streets, regional roads and many other detailed streets are encompassed in the dataset of Chinese ordnance survey map. The applied dataset of road network in this research is derived from the ordnance survey dataset 2013, which has been compared with online datasets (e.g. Google Map and Baidu Map) to prove its secured accuracy. The employed database in this study is made of 6,026,326 segments with the total length of 2,623,867 km (Figure 3).

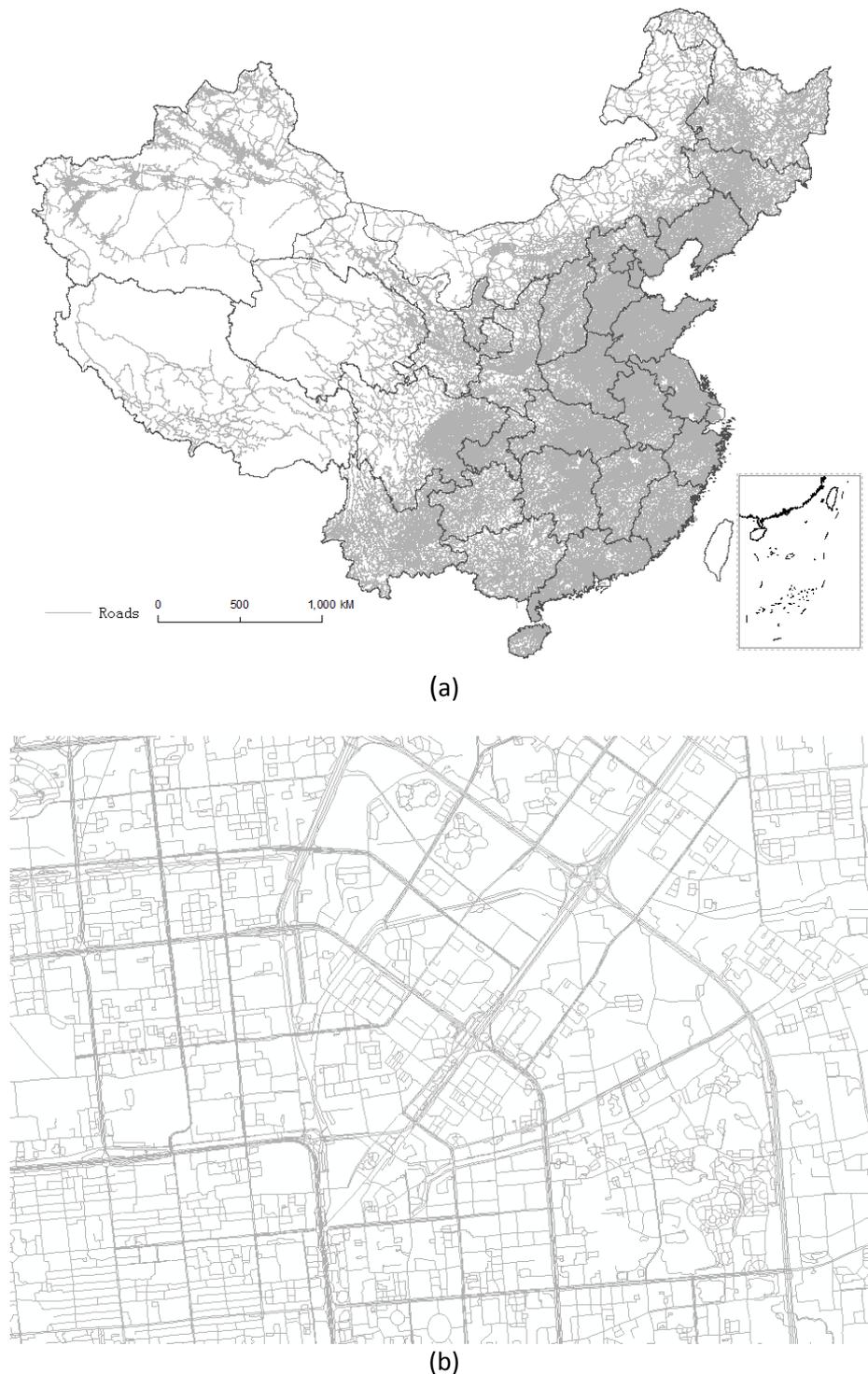

(a)

(b)

Figure 3 Ordnance roads of China in 2012. (a) the whole China, (b) a part of the central city of Beijing

POIs containing 5,281,382 points totally are gathered from business cataloguing websites. There are 20 types of POIs in the initial dataset, and each type refers to one detailed sort of urban activity. All the POIs are adopted in this empirical study to measure the land use density through calculating their amount for each generated parcels.

## 3 METHODOLOGY

### 3.1 The proposed framework

Generally, the empirical framework for delineating urban areas contains three steps based on well propagated data: parcel generation, urban parcel selection (vector CA module) and urban area production (parcel mergence) (Figure 4). In the first step, all possible parcels are defined depending on the fine-scaled road layers in ordnance survey. Then in the following stage, the parcels are inferred with their geometrical and geographical properties and POIs density to automatically select the urban ones in a vector CA approach. Finally, all the urban parcels are dissolved and mapped thereby clearly generating urban areas. All these steps will be illustrated fully in the following sections.

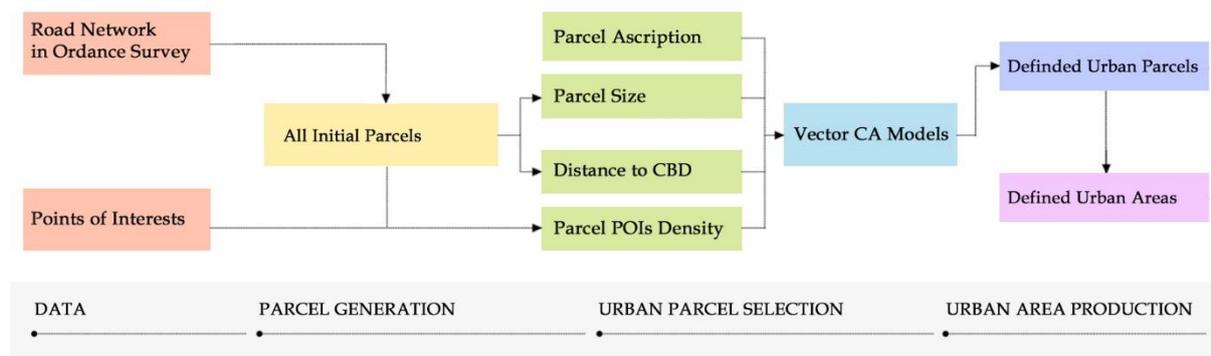

Figure 4 Flow chart of the proposed framework

### 3.2 Generating parcels and inferring their density

Parcels are important spatial units for contemporary urban planning & design and urban studies. In this sub-section, parcels are defined as a continuous built-up area enclosed by roads. Supporting by this idea, all the possible urban parcels are generated by using road layer in ordnance survey. Before generating parcels, the road layers should be processed according to their hierarchy respectively before being merged as a single layer. More specifically, all the segments are connected with tolerance of 20 meters, whereas street segments shorter than 200m are trimmed to avoid cul-de-sacs. Moreover, all the width of all roads is also defined relying on their hierarchy. At last, all initial parcels are presented when the roads are removed from the study areas.

Four properties are further calculated for each parcel. The first two refers to the geometric characteristics of each parcel including the size and compactness determined by the parcels' shape. In addition, the accessibility is also taken into consideration as a locational variable for describing a parcel. Another characteristic is the functional attribute of a parcel for reflecting its actual use. POIs within or close to a parcel are measured as its urban density. Due to the natural unevenness of urban density between central cities and other ones, POIs density is further normalized and placed between 0 and 1 to release the heterogeneity among cities. Because of lacking further attributes of POIs, the popularity of POIs is assumed as the same in this study. When any substitutions are available, they can be expected to approximate the intensity of urban activities explicitly. Hence, in the way of using road network and POIs to identify initial parcels, the spatial and functional features are incorporated together for further urban parcel selection.

## 3.3 Selecting urban parcels by using vector CA

Vector-based constrained cellular automata models are used for picking up urban parcels from the initial ones generated by road network in diverse cities. We suppose this process is similar with that for modeling urban expansion, which sees extensively CA applications. Apart from conventional raster CA model (Batty, et al, 1999), vector-based CA model here depends on irregular polygons rather than regular cells. In this research, each parcel is regarded as a cell with a status that is 0 (urban) or 1 (non-urban). This can be illustrated as a formula as the following.

$$S_{ij}^{t+1} = f(S_{ij}^{t}, \Omega_{ij}^{t}, Con, N)$$

Here, a parcel's status at *t* +1 is considered as a function *f* of parcel's statues and other factors at *t*. In this function, $S_{ij}^{t}$ and $S_{ij}^{t+1}$ denote to the statues of parcels at time points of *t* and *t*+1 respectively; $\Omega_{ij}^{t}$ is the neighboring situation; *Con* refers to the constrains and the *N* is the amount of all parcels. This function can be further transferred to a detailed probability formula:

$$P_{ij}^{t} = (P_{l})_{ij} \times (P_{\Omega})_{ij} \times con(\cdot) \times P_{r}$$

In this function, the possibility of transformation of parcel's state at *t* is illustrated as multiplied product of probabilities of factors. Specifically, $(P_{l})_{ij}$ stands for the local potential that a parcel would convert its status from the non-urban to the urban while $(P_{\Omega})_{ij}$ denotes the conversion possibility in terms of the neighboring situations; $con(\cdot)$ stands for constrains and $P_{r}$ is the stochastic term.

The proposed spatial and functional characteristics are reflected in measuring the local potential. This could be explained in the formula below using a logistic regression model (Wu, 2002):

$$(P_{l})_{ij} = \frac{1}{1+\exp[-(a_{0}+\sum_{k=1}^{m}a_{k}c_{k})]}$$

Where $a_{0}$ is a constant, $a_{k}$ is an estimated coefficient responding to the spatial variable $c_{k}$ and *m* is the total amount of spatial variables. As a result, spatial and functional factors are bonded to reflect parcel's state in this study. Parcel size measured in the natural logarithm of area. Compactness is calculated by the rate that perimeter square subdivided by area. Accessibility is abstracted using the minimum Euclidian distance to the city center. On the other hand, the functional factor is presented by applying the standardized POIs density, which is measured through calculating the rate of raw density in the max density in the samples.

The neighboring potential for a parcel is measured by the amount of peripheral urban parcels around it. This can be defined as:

$$(P_W)_{ij} = \frac{\sum con(S_{ij}^t = urban)}{n}$$

For parcel *ij*, $con(S_{ij}^t = urban)$ stands for the urban parcels within fix areas while *n* is the sum of all accessible parcels. The adjacent relation is defined as 500 m around the parcel *ij*.

Two layers - the steep area (a slope over 25 degrees) and various water bodies, are included as the restrictive condition. Urban expansion is forbidden in these areas. The constraints are expressed as $con(cell_{ij}^t = suitable)$ with a value of 0 or 1, where 1 indicates that there is no restriction on the parcel's development as urban while 0 indicates that the parcel is forbidden to be urban.

The stochastic disturbance $P_r$ in the model stands for any possible change of local policies and accidental errors. It is calculated using

$$P_r = 1 + (-\ln \gamma)^\beta$$

where $\gamma$ is a random number ranging from 0 to 1, and $\beta$, ranging from 0 to 10, controls the effect of the stochastic factor.

Furthermore, by comparing the measured probability $(P_l)_{ij}$ with a fixed threshold value $P_{thd}$, the parcel's status at *t*+1 could be detected. If the measured value is greater than the threshold, the parcel is considered to be urban, if not, the parcel will stay as non-urban. This progress can also be presented as a binary expressions:

$$S_{ij}^{t+1} = \begin{cases} Urban \text{ for } P_{ij}^t \succ P_{thd} \\ NonUrban \text{ for } P_{ij}^t \leq P_{thd} \end{cases} \qquad \sum_{k=1}^{m} A_{ij} \leq A_{total}$$

Finally, for controlling the total area of all urban parcels, the statistics of urban area in 2012 for each city are applied as the upper limits for the total area of selected urban parcels. $A_{ij}$ here is the area of parcel *ij*, and $A_{total}$ denotes to the reported total area for the city in MOHURD (2013).

### 3.4 Mapping urban areas using selected urban parcels

We need translate selected urban parcels into urban areas, considering that street spaces and small unselected urban parcels surrounded by urban ones are also included in urban areas in planning practices. In order to map the urban areas of all cities in China, the selected urban parcels are re-merged into the integrated urban lands in ESRI ArcGIS using the toolbox Aggregate Polygons to present the urban areas for each city (see Figure 5). This tool is used for moderate scale reduction and aggregation on selected urban parcels. Aggregation will only happen where two parcels are within the specified aggregation distance to each other. According to the facts of Chinese urban parcels, the distance to be satisfied between parcel boundaries for aggregation to happen is set 500 m and the minimum area for an aggregated parcel to be retained is 1 ha. In addition, orthogonally shaped output urban areas are created for preserving the geometric characteristic of anthropogenic urban parcels. The projected approach is conducted in all 654 reported cities individually to speed up the parcel aggregation process. Urban areas of each city can then be mapped based on selected urban parcels.

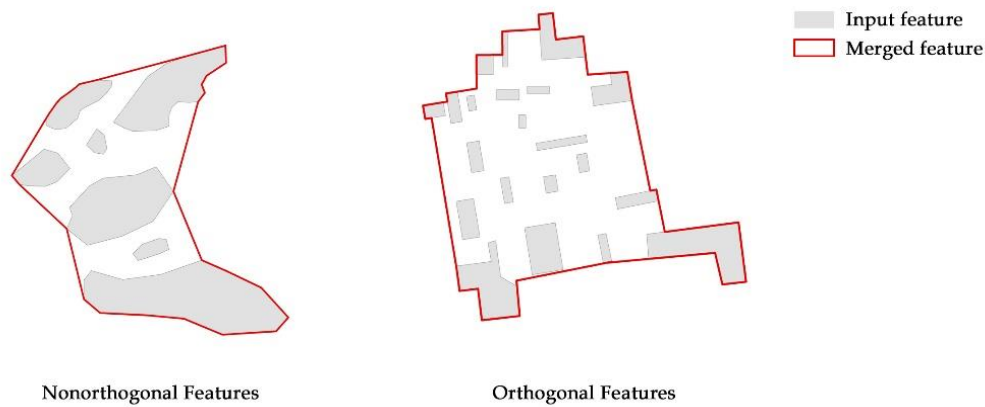

Figure 5 Illustration on converting urban parcels to urban areas (combining polygons within a specified distance of each other into new polygons).

### 3.5 Model validation

For well validating our proposed model, self-validation and external justification are progressed separately and cross-verified the applicability in delineating urban areas for various cities. On the internal dimension, all cities are ranked to detect the scaling law of city size horizontally and on the other hand, urban parcels within typical cities at various administrative levels are also rated for finding the linear relation on a logarithmic scale. In addition, the automatic generated result of urban areas for all the Chinese cities is compared with the outcomes produced by several classical methods reviewed in the introduction section based on geographical coverage datasets containing remote sensing images, census-based population density and road intersection density.

### 3.6 Constrained inversion

A whole process of automatic identification of urban areas could be described as 'holistic constrained inversion' of urban areas (Figure 6), which is a method for speculating very large and complex urban areas based on a relatively small amount of observed data. To avoid subjective factors setting, this model requires the verified effective constraints/parameters in some typically observed cases at the first stage. This process could be thought to be a 'partial inversion', a way to generate key constraints in defining urban areas. In this study, the urban density measured by POIs density is proven to be a key factors in identifying urban parcels in each city. In so-called 'holistic constrained inversion', there are three steps: the first part is about parcels segmentation while the second part is for parcels' mergence. These two steps interact with each other by using POIs density and other factors to select urban parcels as shown above. This proposed open framework has several potentials in addressing the questions of identifying urban areas: first, it offers a method to generate parcel-based urban areas in a large scaled manner relying on universal rules discovered in typical samples; moreover, it can be used as a reference to validate the surveyed urban areas; last but not the least, it could further implies the potential role of omnipresent dataset on duplicating urban areas from a vast scope. Therefore, instead of only promoting our model of delineating urban areas based on POIs and road network, we are also advocating an open framework for presenting 'parcel-up' distributions of urban areas by combining holistic and partial inversions at different scales. That is to say, the model discussed in this paper is an open system combining local equation-based analysis and global simulation, which is ensured for future development in the background that location-based data is increasingly available.

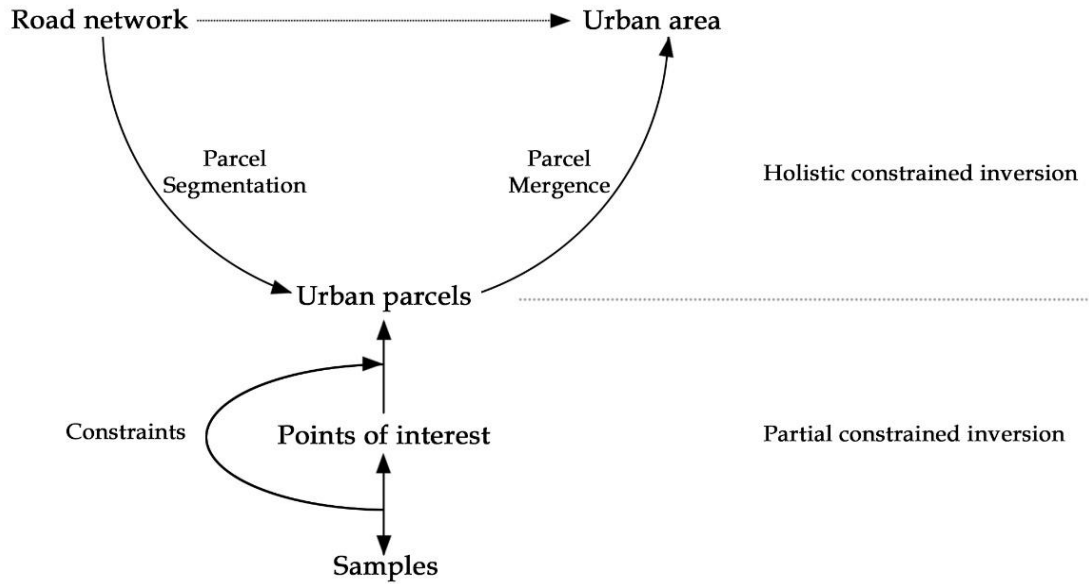

Figure 6 Constraints inversion of identifying urban areas

## 4 RESULTS

### 4.1 Model calibration for vector CA model

Logistic regression is conducted for calibrating the weights for constraints in the proposed vector CA models. But due to the data availability, it is nearly impossible to calculate the weights of controlling factors for each city thereby reflecting the spatial heterogeneity between cities. Hence, the 2010 parcels dataset in Beijing City manually prepared by urban planners in BICP is applied as a typical example of all the other cities. It covers an area of totally 12,183 $km^2$ at a very fine-scaled urban parcel scale (Yanqing and Miyun counties in the Beijing Metropolitan Area are excluded from Beijing City). There are totally 52,330 parcels reported, among which 36,914 parcels are identified as urban ones.

According to the result of binary logistic regression (Table 1), 78.9% of all parcels can be explained by the generated function. And all factors except compactness have passed $p$ test, revealing that they are significantly related to the differences between non-urban and urban ones. This logistic regression results have been employed in vector CA models for all the Chinese cities. In order to test the accuracy of this model, the generated results of Beijing City by the CA model was compared with the BICP dataset again, and then an overall explanatory ability of 81.5% indicates that the applicability of our model in delineating urban areas in terms of urban parcels.

Table 1 Binary logistic regression results for BICP parcels

| Name | Coefficient | S.E, | Sig. |
|---|---|---|---|
| Constant | 5.359 | .058 | 0.000 |
| Natural logarithm of parcel size | -0.306 | .006 | 0.000 |
| Distance to the city center | -0.099 | .001 | 0.000 |
| POIs density | 3.431 | .085 | 0.000 |

## 4.2 Selected urban parcels

The proposed constrained vector CA model was conducted in all 654 reported cities in China, in which a sum of 707,330 urban parcels with 51,286 km$^2$ in area were detected and labeled as 'urban' among all 851,054 initial parcels (Figure 7). The average numbers of urban parcels in cities on various administrative levels differentiate with each other significantly. Precisely, there are 1411 urban parcels in MD averagely, followed by 407 in SPC, 199 in OPCC, 79 in PLC and 26 in CLC respectively. When scrutinizing these statistics, the more population or higher administrative ranks the cities occupied, the greater number of urban parcels they will have. In other words, the scaling laws of the population or city size also can be significantly observed in terms of the amount of urban parcels in each city.

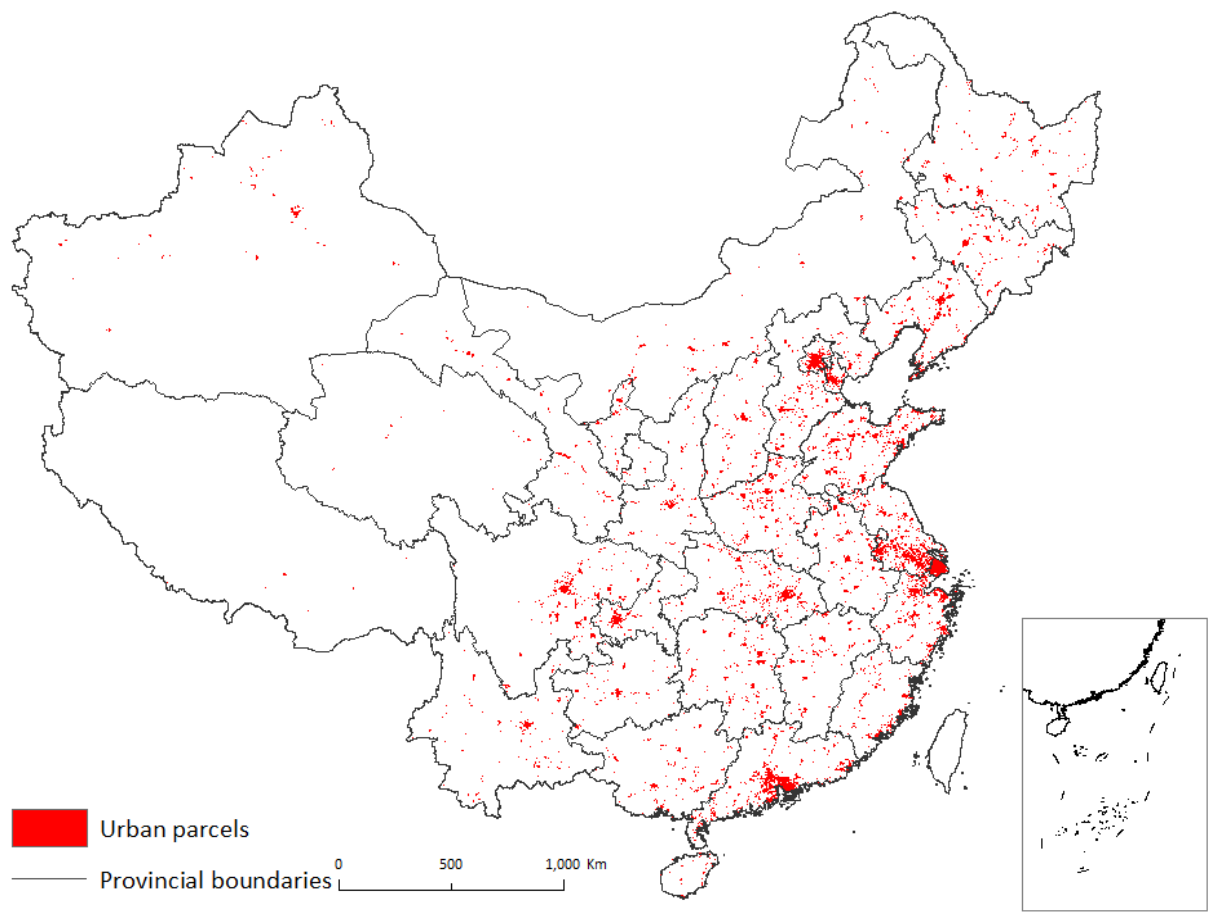

Figure 7 Selected urban parcels for all Chinese cities

Scaling law is a universal rule not only for natural phenomenon but also for urban areas. On a logarithmic scale, this relationship between the size of urban areas and their frequency distribution should be linear. It also enable to self-validate the proposed vector CA models for each city in this paper (Vilet et al., 2009). In order to verify the performance between cities, the size (in term of the number of urban parcels in each city) of all cities is firstly plotted against their ranks (figure 8a). A shape of long tail distribution can be evidently recognized which reveals that there are far more cities with fewer urban parcels than those with a large number of urban parcels (Jiang, 2013). When it comes to the log-log distribution, a perfect power law fit ($R^2$ is 0.988 and alpha is 2.06) can be visibly presented by considering the cities with more urban parcels than the average level. Thus, the significant rank-size pattern with high R-square values indicate the applicability of our models for all cities.

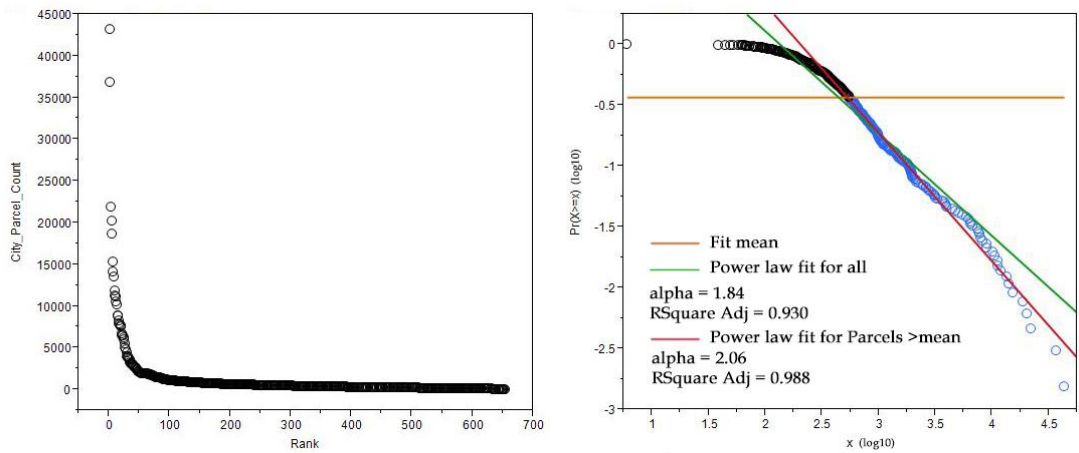

(a)

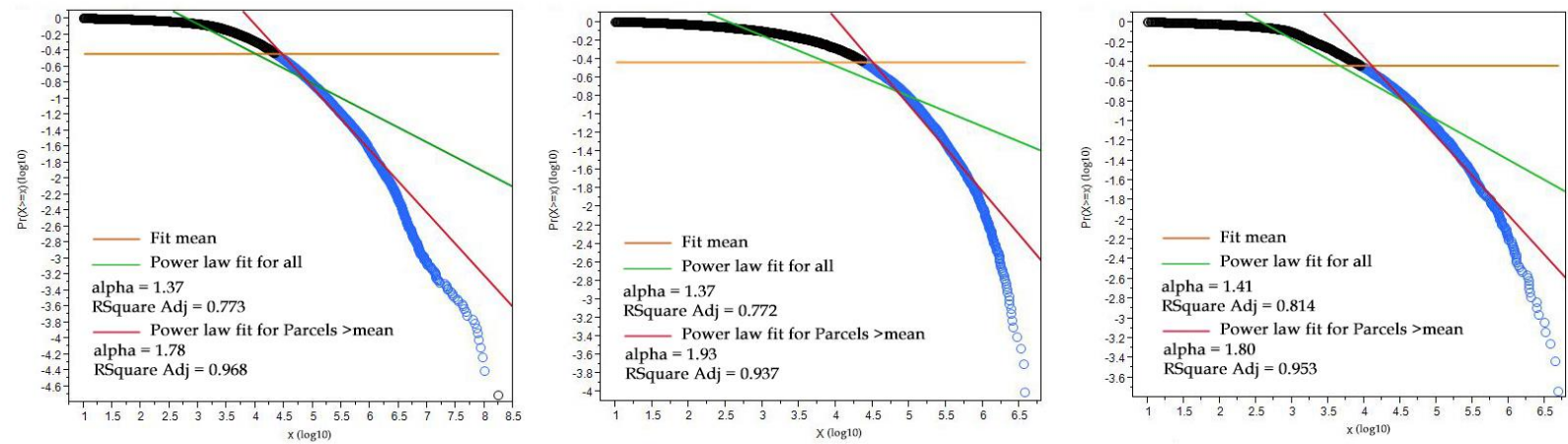

(b)

Figure 8 Power law distributions a) in term of the parcel numbers for all 654 cities; b) in terms of parcel size for typical cities

On the other hand, the power law fit is also adopted in analyzing the ranks of urban parcels' size in typical cities at various governmental levels for understanding the applicability of proposed approach in each kind of cities internally. Generally, the power law fits for all parcels can explained around 70% of urban areas. Better regressions are implied in the cities occupied higher administrative levels rather than ones in the cities at lower levels. More specifically, the alpha value in Beijing, Nanjing and Changsha are all above 1.37 and the adjust $R^2$ is greater than 0.77, whereas the Weifang and Gongzhuling have the alpha value of 1.28 and a smaller $R^2$ which is about 0.65. While removing the parcels less than the mean size, better power law fits can be easily recognized (all R2 increased above 0.9). All these shared similar trends emerged in typical examples suggest that our models can be applied in modeling the urban parcels within different kinds of cities.

### 4.3 Urban areas of all Chinese cities

By merging all the selected urban parcels, urban areas of all Chinese cities are automatically illustrated. For gaining more insights to these results, the typical cities (e.g. Beijing for MD, Nanjing for SPC, Changsha for OPCC, Weifang for PLC and Gongzhuling for CLC) on different administrative levels are listed and compared with the results from other datasets including DMSP/OLS, census-based population density and road intersections (Table 2).

Compared with the results from other datasets, the urban area generated through our approach has generally higher resolution than other databases (Table 2). The outputs captured by using the other three datasets are highly correlated to the ones detected in our projected framework in relatively developed cities in terms of initial eyes judgment. This may be mainly on the basis of good digital infrastructures and small censuses setting for survey. This assumption can be verified through comparison among the results of developing cities. Precisely, our approach seems to be more possible to produce detailed results than the other two. From a more precise point of view, the generated urban area was further overlapped with other results to detect the overlapping rates. More details can be found in model validation (Section 5.1).

Table 2 The profile of urban areas for typical cities

| City | 1 Beijing | 2 Nanjing | 3 Changsha | 4 Weifang | 5 Gongzhuling |
|---|---|---|---|---|---|
| Urban parcels | 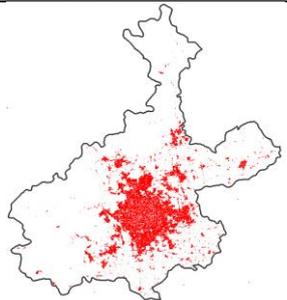 | 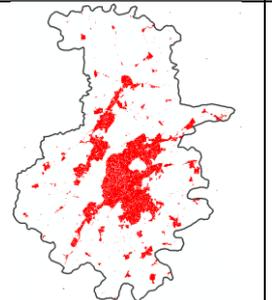 | 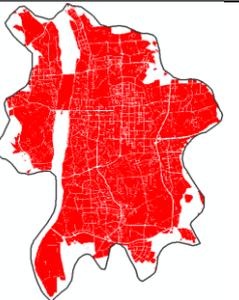 | 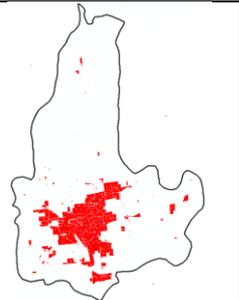 | 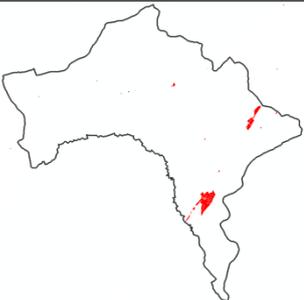 |
| Urban areas | 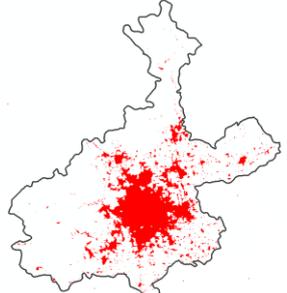 | 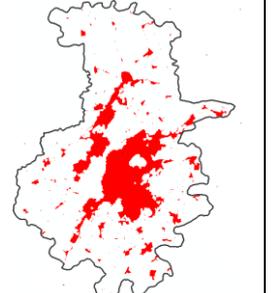 | 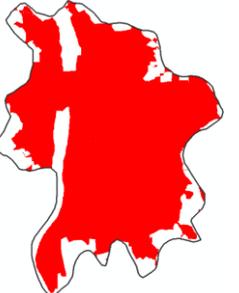 | 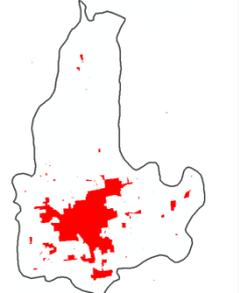 | 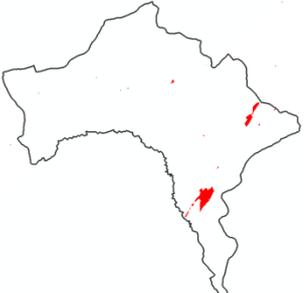 |

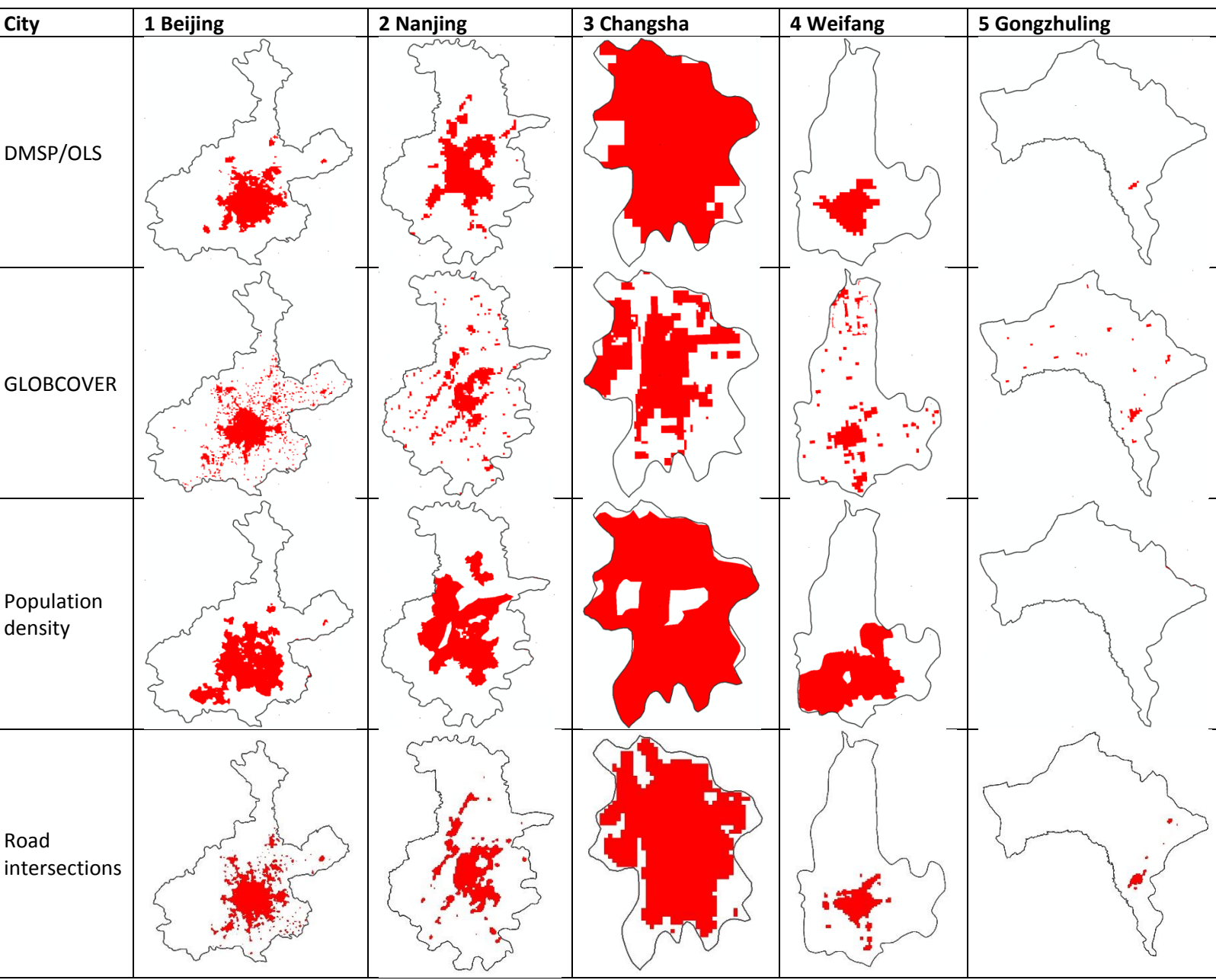

## 5 DISCUSSION

### 5.1 Model validation

Model validation is conducted via comparing with urban areas by our approach with four datasets for all 654 Chinese cities: (1) the urban areas defined in the 300m resolution in GLOBCOVER (Bontemps, 2009); (2) the urban areas presented in the 1km resolution retrieved from DMSP/OLS in 2008 (Yang et al, 2013); (3) the urban areas represented by sub-districts with population DENSITY greater than the mean density (977 people per km$^2$) of all 39,007 sub-districts of China in 2010 by using Jiang (2013)'s head-tail division rule and the 2010 population census of China (Wu et al, 2014); (4) the urban areas presented by road INTERSECTION density using the ordnance survey applied in this study. Urban areas are selected in each city by sorting all grids' estimated kernel density of road intersections while considering observed total area of a city (elaborated in Section 2.2).

All the results are shown in Table 3 for detail representation. In terms of the captured size of urban blocks, the urban parcels in this study (average size is about 300m * 400m) is far smaller than the ones reported in other four datasets, reflecting a fact that better scaled outputs are achieved through our approach. (1) From the perspective of overlapping rate, there are 65.5% of detected common urban areas (totally 30,606 km$^{2)}$ in our outputs intersected with DMSP/OLS. With consideration of time mismatch between these two dataset, our suggested approach can be reasonably imagined to produce good results for all the Chinese cities depending on the evaluation above. (2) The results between proposed data GLOBCOVER are not as good as expected initially. There are only 20,801 km$^2$ urban areas occupied about 44.5% in our result that are intersected with GLOBCOVER. These might be the results of non-correspondence between these two datasets regarding with time and resolution. (3) There are 81.9% of urban areas by our method fall into the urban areas represented by population density, indicating that most of our results are associated with high population density. The overlapped ratio over 80% is partially due to the overestimated urban areas in DENSITY, which is nearly three times of that stated by MOHURD (2013). (4) The comparison results between ours and by using road intersections are acceptable (76.8%), which could be attributed to the same data source used in both methods.

Table 3 The comparison of urban areas in various datasets for 654 cities in China

| Data | Year | Spatial resolution | Urban area (km$^2$) | # patches | Average patch size (ha) | Intersected with ORDNANCE (km$^2$) |
|---|---|---|---|---|---|---|
| ORDNANCE | 2012 | 269 m | 46,713 | 18,404 | 312.5 | N/A |
| DMSP/OLS | 2008 | 300 m | 45,834 | 1,345 | 3,407.7 | 30,606 (65.5%) |
| GLOBCOVER | 2009 | 1 km | 39,789 | 12,701 | 313.3 | 20,801 (44.5%) |
| DENSITY | 2010 | 6.7 km | 126,860 | 728 | 17,425.8 | 38,245 (81.9%) |
| INTERSECTION | 2012 | 500 m | 46,703 | 4221 | 1,106.5 | 35,868 (76.8%) |

Note: All datasets are clipped using the administrative boundaries of 654 cities in China. ORDNANCE stands for the urban areas by this study. Spatial resolution of ORDNANCE is for the average parcel size of all urban parcels (supposed to be square). Spatial resolution of DENSITY is for the average sub-district size among all urban sub-districts in China (supposed to be square).

Admittedly, there is a blank among all existing methods to produce an authoritative urban area. In other words, it is hard to determine which result could most accurate since that each method could reflect one kind of possible understanding of the same question. However, the overall precisions are acceptable taking the differences between various models into account.

### 5.2 Horizontal evaluation on methods for delineating urban areas

In addition to quantitatively comparing our results with methods, there are six dimensions are considered for qualitatively evaluating the strengths of existing methods including practicality, geographical scale, result resolution, data availability, methodological convenience and dynamics (Figure 9). 10 professionals in different planning institutes in China were interviewed and asked to rate the performance of every approach according to their working experience. The practicality here refers to the value that can directly benefit current urban planning. Due to the similarly basic spatial units setting, our urban parcels-based results are generally considered as the most straightforward method to reflect real developments in urban parcels. In the meantime, conventional approaches e.g. DMSP/OLS, survey parcel map and population density maps are also labeled as the practical tools to understand urban extent. Moreover, the proposed approach in this study is expected to better balance the dilemma between the coved scale and result resolution in traditional models (Long, 2014). Relying on the open datasets in the background of development of volunteered geographic dataset, this method projected in this paper is also well-thought-out to be a public accessible and temporally updatable data for urban planning and studies. Regarding with the methodological convenience, spatial survey and statistics based methods are straightly understandable, whereas our approach is understood as direct way of packaging complex simulations in an automatic manner. Frankly, all these evaluations are generally based on the reality of urban planning in developing countries, particularly in China, which means that the assessment addressing the same issue would be different in those developed nations where urban survey and statistics are conducted maturely for many years. However, it is still very worth to promote our produced method in web 2.0. It is a big model (Long, 2014) in a direct, fine-scaled and dynamic sense based on omnipresent open data thereby benefiting the understanding of urban area towards city management and planning.

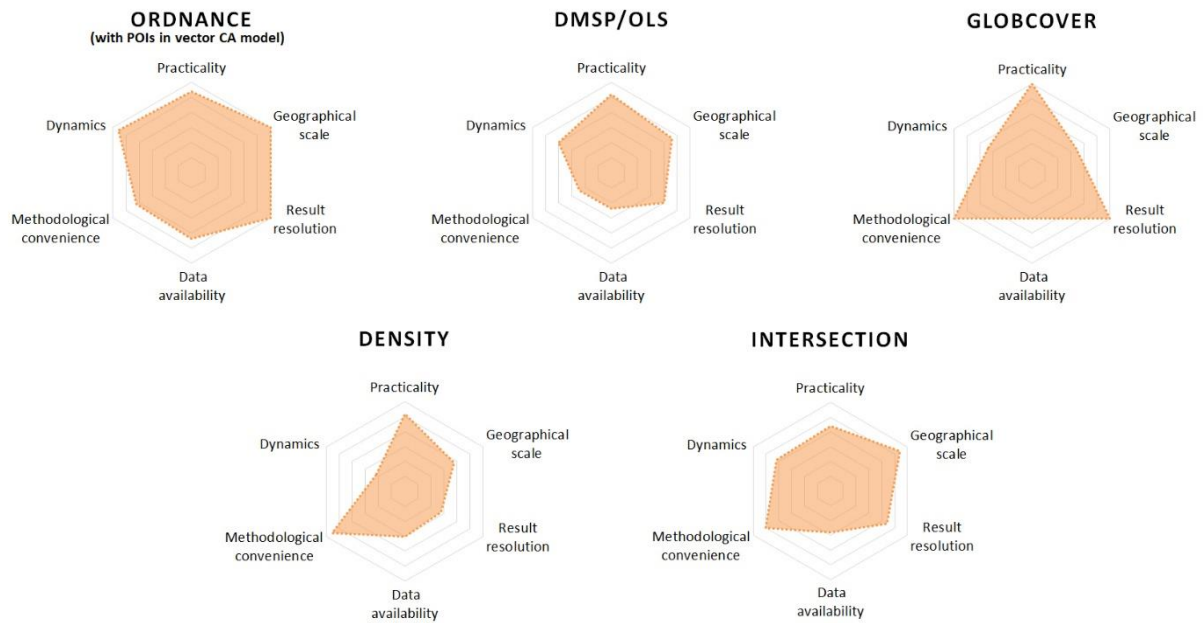

Figure 9 Comparison between existing methods of delineating urban areas

## 5.3 Potential bias and further steps

This study proposes an automatic framework to generate urban area and provides examples of all Chinese cities. The increasingly available volunteer geographic information in this framework also promotes the merits of this approach. Nevertheless, several limitations still exist in the current study, which would be highlighted in our future research. **First**, current methodology could be directly improved based on the increase of open data availability. The location based online information (e.g. check-ins) could be considered to infer the weights of POIs thereby reflecting the actual urban usage more accurately. **Second**, more samples of cities should be used for model calibration to enhance precision of our approach, even if our methods have already proven the applicability and flexibility of applying local constraints rather than the global ones for all cities. **Last**, the presently applied city-level urban area statistics could be replaced by fined scaled ones, e.g. districts or even sub-districts, aiming for controlling the total area of urban parcels in a more detailed manner.

# 6 CONCLUDING REMARKS

In this paper, a vector cellular automata model is proposed based on road network in ordnance survey and points of interest for explicitly delineating parcel-based urban areas. Urban areas in all 654 cities are generated by using our approach. The whole progress contains several components including parcels generation, urban parcels selection, and urban area production. In the first step, road network layer in ordnance survey is applied to define parcels by removing the buffered roads from the study area. In the following stage, all the parcels are equipped with their attributes like size, compactness, accessibility, and POIs density. And then, the vector cellular automata model is adopted for identifying urban parcels from all generated parcels, taking into account spatial variables of each parcel as well as conveyed total area in each city. At last, the urban areas of each city are mapped by aggregating urban parcels. In the process of self-validation, the power law fits are detected when analyzing the relationships between generated amounts of urban parcels and their ranks across cities and the correlations between parcels' size and the frequency distributions in five typical cities, proving the applicability of our approach. The final results are also validated through comparing them with urban areas presented by DMSP/OLS, GLOBCOVER, population density and road intersection density maps. Furthermore, after interviewing relevant urban planners, the proposed approach in this paper is given high ranking on various dimensions. In sum, our model is proven to be not only effective in modeling urban area through incorporating spatial and functional features of urban parcels but also more straight-forward, time-saving and fine-scaled, compared with other existing ones.

As an alternative approach to capturing urban areas for all the Chinese cities partially based on open data, the presented method is expected to have contributions on several aspects. **At the outset**, our study for delineating urban areas of all Chinese cities from a national perspective in a fine-scaled manner would fill a gap in existing literature. At the same time, the produced approach could promote urban studies of urban areas based on open dataset and enhance the relevant studies in web 2.0. **Furthermore**, this method could be regarded as a useful presentation and even solution for the cities in developing countries, particularly the small cities limited by the digital infrastructures. **Finally**, due to well-consideration of omnipresent spatial and functional factors, the introduced method could reclaim the need for consistency, efficiency and availability in defining city urban areas across cities.

The projected framework has potentials to benefit relevant urban studies and policy distributions. **First**, through this study, the current situation of urban development can be reflected at a standard level, thus feeding both intra-city and inner-city academic studies. It would be more helpful for relatively small cities where digital infrastructures are poor and fine-level statistics are hardly secured. On the other hand, it can significantly save the cost for collecting data temporally without too much investment. This could be approached on our releasing the urban areas delineated on acceptance of this manuscript. **Second**, urban area simulation process could promote a deeper understanding of the 'parcel-up' urbanism that reflects the phase of a large site divided and sold off for development. **Third**, this model can be further developed to an advanced one to simulate the urban expansion. It might directly benefit urban planning prediction and evaluate the effectiveness of strategies and policies. **Fourth,** our model can methodologically helpful in unifying the calibers of defining urban areas amongst diverse cities based on ubiquitous data and reclaiming the need for consistency, efficiency and availability in defining urban areas with well-consideration of omnipresent spatial and functional factors across cities.